\documentclass[showpacs,twocolumn,showkeys,amsmath,amssymb,floatfix,superscriptaddress,aps,prl]{revtex4}
\usepackage{graphicx}
\usepackage{amssymb}
\usepackage{amsmath}
\usepackage{dcolumn}
\usepackage{bm}
\include{natbib}

\begin{document}

\newtheorem{theo}{Theorem}

\title{On the Limiting Cases of Nonextensive Thermostatistics}
\author{Michele Campisi}
\email{campisi@unt.edu} \affiliation{Department of Physics,
University of North Texas, P.O. Box 311427, Denton, TX 76203-1427,
USA}
\date{\today}

\begin{abstract}
We investigate the limiting cases of Tsallis statistics. The
viewpoint adopted is not the standard information-theoretic one,
where one derives the distribution from a given measure of
information. Instead the mechanical approach recently proposed in
[M. Campisi, G.B. Bagci, Phys. Lett. A (2006),
doi:10.1016/j.physleta.2006.09.081], is adopted, where the
distribution is given and one looks for the associated physical
entropy. We show that, not only the canonical ensemble is
recovered in the limit of $q$ tending to one, as one expects, but
also the microcanonical ensemble is recovered in the limit of $q$
tending to minus infinity. The physical entropy associated with
Tsallis ensemble recovers the microcanonical entropy as well and
we note that the microcanonical equipartition theorem is recovered
too. We are so led to interpret the extensivity parameter q as a
measure of the thermal bath heat capacity: $q=1$ (i.e. canonical)
corresponds to an infinite bath (thermalised case, temperature is
fixed), $q=-\infty$ (microcanonical) corresponds to a bath with
null heat capacity (isolated case, energy is fixed), intermediate
$q's$ (i.e. Tsallis) correspond to the realistic cases of finite
heat capacity (both temperature and energy fluctuate).
\end{abstract}

\keywords{orthode, microcanonical, canonical, finite heat bath,
Tsallis, R\'{e}nyi} \pacs{05.20.-y; 05.30.-d; 05.70. ; 03.65.-w}

 \maketitle

It is a very well known fact that nonextensive thermostatistics,
based on q-exponential power-law ensembles generalizes the
standard statistical mechanics stemming from the canonical
ensemble, to which it tends as the non-extensive parameter $q$
approaches 1. This generalization has been first proposed within
an information-theoretic approach where the standard Shannon-Gibbs
entropy is replaced by the order-q Tsallis entropy. As $q$
approaches 1, the Tsallis entropy approaches Shannon-Gibbs entropy
and one is able to recover standard statistical mechanics
\cite{Tsallis88}. Very recently a new approach has been proposed
to study the theoretical foundations of nonextensive
thermostatistics \cite{Campisi06}. This approach is based on the
concept of ``orthode'' originally proposed by Boltzmann in the
1880s and is completely independent from the information-theoretic
one \cite{Gallavotti}. The approach provides a criterion to assess
whether a given statistical ensemble provides a ``mechanical model
of thermodynamics'', and, if this is the case, tells what is the
physical entropy associated to a given ensemble. In this respect
the method proceeds along the opposite path followed in the
information-theoretic approach where, instead, one starts from an
entropy function and derives the ensemble. In a previous work
\cite{Campisi06} we have shown that the Tsallis ensemble:
\begin{equation}\label{eq:Tsallis-ensemble}
    \rho_q (\textbf{z};U,V) = \frac{\left[ 1 -
\beta(1-q)(H(\textbf{z};V)-U)\right]^{\frac{q}{1-q}}}{N_q},
\end{equation}
where
\begin{equation}\label{eq:EN}
    N_q(U,V) = \int d\textbf{z} \left[ 1 -
(1-q)\beta(H(\textbf{z};V)-U)\right]^{\frac{q}{1-q}},
\end{equation}
is an ``orthode'', namely it provides a \emph{mechanical model of
thermodynamics} in the sense mentioned above. Within such a
mechanical approach the entropy is given by the $1/q$-order
R\'{e}nyi information associated with the distribution $\rho_q$ :
\begin{equation}\label{eq:S=logEN}
    S^{[q]}(U,V) = \log
     \int d\textbf{z} \left[ 1 -
(1-q)\beta(H(\textbf{z};V)-U)\right]^{\frac{1}{1-q}}
\end{equation}
This entropy is consistent with the maximum information-entropy
principle \cite{Campisi06,Bashkirov04} and can be considered as a
good expression for the \emph{physical entropy} \cite{Campisi06}.
As we shall see, when $q$ tends to 1, the distribution
(\ref{eq:Tsallis-ensemble}) tends to the canonical distribution
\emph{and} the entropy (\ref{eq:S=logEN}) tends to the canonical
entropy. In sum the Tsallis ensemble provides a \emph{mechanical}
generalization of canonical ensemble. The main result of the
present work is that the same ensemble can be seen as a
\emph{generalization of the microcanonical ensemble} as well, to
which it reduces when the index $q$ tends to minus infinity. We
shall also provide a physical interpretation for the
generalization scheme where canonical and microcanonical ensemble
represent two limiting cases.

According to Boltzmann's reasoning, one family (i.e., ensemble) of
distributions, $\rho(\mathbf{z};\lambda_i)$ parameterized by a
given number of parameters $\lambda_i$, provides a mechanical
model of thermodynamics, i.e., it is an orthode if, defined the
macroscopic state of the system by the set of following
quantities:
\begin{equation}{\label{eq:statedef}}
\begin{tabular}{l}
$U_\rho= \left\langle H \right\rangle_\rho $\quad``energy'' \\
$T_\rho= \frac{2\left\langle K\right\rangle_\rho}{n}$\quad
``doubled kinetic energy per degree of freedom''\\
$V_\rho= \left\langle V \right\rangle_\rho $ \quad ``generalized displacement'' \\
$P_\rho=\left\langle -\frac{\partial H}{\partial
V}\right\rangle_\rho$ \quad ``generalized conjugated force'',
\end{tabular}  \end{equation}
for infinitesimal and independent changes of the $\lambda_i$'s,
the \emph{heat theorem}
\begin{equation}\label{}
    \frac{dU_\rho+P_\rho dV_\rho}{T_\rho} = exact\quad differential
\end{equation}
holds \cite{Gallavotti,Campisi05}. In sum the ensemble provides a
mechanical model of thermodynamics if the average of certain
mechanical quantities evaluated over the ensemble's distributions
are related to each other according to the fundamental equation of
thermodynamics. If this is the case, the quantity that generates the
differential is to be interpreted as the physical entropy associated
with the given ensemble.

For example the canonical ensemble, which is parameterized by
$(\beta,V)$, reads:
\begin{equation}\label{eq:canonical-ensemble}
\rho_c(\mathbf{z};\beta,V)= \frac{e^{-\beta
H(\mathbf{z};V)}}{Z(\beta,V)}
\end{equation}
where
\begin{equation}\label{}
Z(\beta,V) = \int d\textbf{z} e^{-\beta H(\mathbf{z};V)}
\end{equation}
The corresponding entropy is given by the well known formula:
\begin{equation}\label{Eq:canonicalEntropy}
    S_c(\beta,V) = \beta U_c(\beta,V) + \log Z (\beta,V)
\end{equation}
To see that this function generates the heat differential, it is
enough to take the partial derivatives of $S_c$ and use the
canonical equipartition theorem:
\begin{equation}\label{eq:equiTeo-canonical}
   T_c = \left\langle p_i\frac{\partial H}{\partial p_i} \right\rangle_c = \frac{1}{\beta}
\end{equation}
where the symbol $<\cdot>_c$ denotes average over the canonical
distribution (\ref{eq:canonical-ensemble}).

Similarly the microcanonical ensemble, which is parameterized by
$(U,V)$, reads:
\begin{equation}\label{eq:microcanonical-ensemble}
\rho_{mc}(\mathbf{z};U,V)=
\frac{\delta(U-H(\mathbf{z};V))}{\Omega(U,V)}
\end{equation}
where
\begin{equation}\label{eq:Omega}
\Omega(U,V) = \int d\mathbf{z}\delta(U-H(\mathbf{z};V))
\end{equation}
denotes the structure function \cite{Khinchin}. As shown in Ref.
\cite{Campisi05} the corresponding entropy is given by
\begin{equation}\label{}
    S_{mc}(U,V) = \log \int d\mathbf{z}\theta(U-H(\mathbf{z};V))
\end{equation}
To see that this is the right entropic function it is enough to
take the partial derivatives and use the microcanonical
equipartition theorem \cite{Khinchin}:
\begin{equation}\label{eq:equiTeo-microcanonical}
    T_{mc} = \left\langle p_i\frac{\partial H}{\partial p_i} \right\rangle_{mc} = \frac{\Omega(U,V)}{\Phi(U,V)}
\end{equation}
where $<\cdot>_{mc}$ denotes average over the microcanonical
distribution (\ref{eq:microcanonical-ensemble}) and
\begin{equation}\label{}
\Phi(U,V) = \int d\mathbf{z}\theta(U-H(\mathbf{z};V)).
\end{equation}

As the work of Ref. \cite{Campisi06} has highlighted, a completely
similar situation occurs for the ensemble
(\ref{eq:Tsallis-ensemble}), and the entropy (\ref{eq:S=logEN}).
Namely the work of Ref. \cite{Campisi06} has shown that the
 Tsallis ensemble of Eq. (\ref{eq:Tsallis-ensemble}) is
an orthode too. The same work also suggested that the Tsallis
ensemble can be considered as a \emph{hybrid} ensemble. This
refers to the fact that both $\beta$ and $U$, appear explicitly in
its expression, therefore one can choose between two possible
parameterizations: either fixes $U$ and adjusts $\beta$ in such a
way that
\begin{equation}\label{eq:U=<H>}
   U= \left\langle H \right\rangle,
\end{equation}
or, fixes $\beta$ and adjusts $U$, in such a way that the same
equation be satisfied (throughout this paper, the symbol $<\cdot>$
denotes average over Tsallis ensemble
(\ref{eq:Tsallis-ensemble})). In the first case one has $\beta=
\beta(U,V)$, and the ensemble will be parameterized by $(U,V)$, in
the second $U=U(\beta,V)$, and the parameters will be $(\beta,V)$.
In sum, there is a \emph{duality}, between two possible
representations: the microcanonical-like one and the
canonical-like one. So, according to the parametrization adopted,
one can see the Tsallis ensemble either as a generalized canonical
ensemble or as a generalized microcanonical ensemble. This is true
not only from a qualitative viewpoint, but also from a
quantitative one. As it is well known, the ensemble
(\ref{eq:Tsallis-ensemble}) is indeed a generalization of the
canonical one to which it tends when $q$ goes to $1$. In fact, for
the well-known properties of the $q$-exponential \cite{Tsallis98},
we have:
\begin{equation}\label{eq:lim-q-1-rho}
\lim_{q \rightarrow 1} \rho_q =\frac{e^{-\beta H}}{Z} = \rho_c,
\qquad \lim_{q \rightarrow 1}S^{[q]} = \beta U + \log Z = S_{c}
\end{equation}

We would like to stress that, in this limit the explicit
dependence of the distribution $\rho_q$ on $U$ disappears. Namely
the \emph{duality} property is lost in the limit $q \rightarrow
1$, as the only possible parametrization is the $(\beta,V)$ one.

\begin{figure}
  \includegraphics[width=8cm]{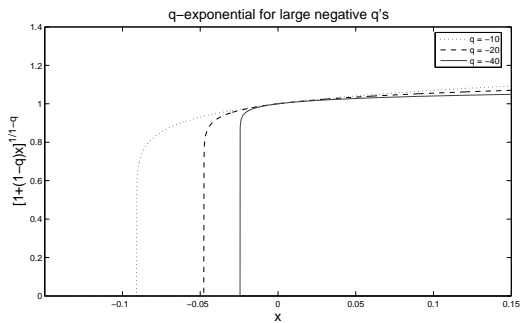}
  \caption{Plot of  q-exponential (Eq. \ref{eq:Eq}) for large negative values of $q$. The function tends to the Heaviside step function.}\label{fig:tsallis2}
\end{figure}

\begin{figure}
  \includegraphics[width=8cm]{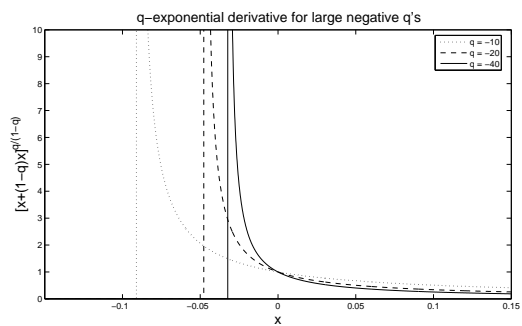}
  \caption{Plot of q-exponential derivative (Eq. \ref{eq:E'q}) for large negative values of $q$. The function approaches a Dirac
  delta function.}\label{fig:tsallis}
\end{figure}

In a similar manner the microcanonical orthode is a special case
of Tsallis orthode recovered in the limit $q \rightarrow -\infty$,
in which case the explicit dependence on $\beta$ disappears. In
this limit, again, the \emph{duality} is lost and the only
possible parametrization is the $(U,V)$ one. Considering the
q-exponential function:
\begin{equation}\label{eq:Eq}
    e_q(x)=\left\{\begin{tabular}{ll}
      0 &  $1+(1-q)x < 0$\\
      $\left[1+(1-q)x \right]^{1/(1-q)}$ & $1+(1-q)x \geq 0$ \\
    \end{tabular} \right.
\end{equation}
it is easily seen that:
\begin{equation}\label{}
    \lim_{q \rightarrow -\infty} e_q(x) =\left\{\begin{tabular}{ll}
      0 &  $x < 0$\\
      1 & $x \geq 0$ \end{tabular}\right. =
      \theta(x).
\end{equation}
This fact is illustrated in Figure \ref{fig:tsallis2}.
Therefore one has
\begin{equation}\label{}
    \lim_{q \rightarrow -\infty} S^{[q]}
     = S_{mc}
\end{equation}
where we have used the fact that $\theta(\beta x)= \theta(x)$ for
$\beta >0$. Further one has that the Tsallis ensemble
(\ref{eq:Tsallis-ensemble}) is expressed in terms of the
``derivative'' of the q-exponential
\begin{equation}\label{eq:E'q}
    e'_q(x) = \left\{\begin{tabular}{ll}
      0 &  $1+(1-q)x < 0$\\
      $\left[1+(1-q)x \right]^{q/(1-q)}$ & $1+(1-q)x \geq 0$\\
    \end{tabular} \right.
\end{equation}
which tends to a Dirac delta function. This is graphically
illustrated in Fig. \ref{fig:tsallis}.
A simple way to prove this result is to consider the Fourier
transform of the function $e'_q$:
\begin{equation}\label{}
    \widehat{e'_q}(\omega) = e^{-i\omega \alpha_q}\frac{\Gamma(\alpha_q+1)}{(-i\omega \alpha_q)^{\alpha_q}}
\end{equation}
where for simplicity we have used the notation $\alpha_q=1/(1-q)$.
Now it is easily seen that, in the limit $q \rightarrow -\infty$
($\alpha_q \rightarrow 0^+$), $\widehat{e'_q}(\omega) \rightarrow
1$ which is the Fourier transform of the Dirac delta. Therefore
the function $e'_q(x)$ tends to the delta function. From this it
follows that the distribution $\rho_q$ tends to the microcanonical
distribution:
\begin{equation}\label{}
    \lim _{q \rightarrow -\infty} \rho_q 
    = \frac{\delta(U-H(\textbf{z};V))}{\int
    \delta(U-H(\mathbf{z};V))} = \rho_{mc}
\end{equation}
It is evident that, due to the property $\delta (\beta x) =
\beta^{-1} \delta (x)$, in the limit $q \rightarrow -\infty$, the
explicit dependence on $\beta$ disappeared. This is understood
also based on the fact that, being the distribution extremely
peaked, the average of $H$ always equals $U$, no matter the value
taken by $\beta$.

It is important to notice that the microcanonical equipartition
theorem is recovered too. The equipartition theorem associated
with the ensemble (\ref{eq:Tsallis-ensemble}), reads
\cite{Martinez02}:
\begin{equation}\label{eq:T=1/beta-tsallis}
    T^{[q]} = \left\langle p_i\frac{\partial H}{\partial p_i} \right\rangle = \frac{1}{\beta}\frac{N_q}{\mathcal{N}_q}
\end{equation}
where:
\begin{equation}\label{}
    \mathcal{N}_q(U,V) = \int d\textbf{z} \left[ 1 -
(1-q)\beta(H(\textbf{z};V)-U)\right]^{\frac{1}{1-q}}
\end{equation}
For finite q's, thanks to Eq. (\ref{eq:U=<H>}), we have $ N_q =
\mathcal{N}_q$ hence $T^{[q]}= \frac{1}{\beta}$ \cite{Martinez02}.
Therefore the canonical equipartition theorem (see Eq.
\ref{eq:equiTeo-canonical}) is trivially recovered for $q=1$.
Nonetheless, when $q$ goes to infinity, the relation $ N_q =
\mathcal{N}_q$ stops holding. In facts we have
\begin{equation}
\lim_{q \rightarrow - \infty} \mathcal{N}_q = \Phi \nonumber \qquad
\lim_{q \rightarrow - \infty} N_q = \frac{1}{\beta} \Omega
\end{equation}
so that $\lim_{q \rightarrow -\infty} T^{[q]} =
\frac{\Omega}{\Phi}$, namely the microcanonical equipartition
theorem is recovered as well.

Thanks to the \emph{mechanical} approach of Boltzmann, we have
been able to see how the microcanonical and canonical orthodes are
two extremal cases of a more general family of \emph{hybrid}
orthodes, which can be parameterized either through $(U,V)$ or
$(\beta, V)$. We shall refer to this fact as the $\beta
\leftrightarrow U$ \emph{duality}. The microcanonical and
canonical ensembles arise as the two opposite limits where the
duality property is lost.

We shall now investigate the physical meaning of such
\emph{hybrid} statistics. In other words, we shall ask ourselves
in what physical situation we expect to observe Tsallis
statistics. It is well known that the microcanonical ensemble
describes the statistical properties of isolated systems whereas
the canonical one well describes the properties of systems in
contact with a heat bath. Both ensembles apply to two ideal
(nonetheless very useful) cases: that of a system in contact with
a heat bath with infinite capacity (the temperature is fixed) and
that of an isolated system, namely a system in contact with a bath
with null specific heat (the energy is fixed). Between these two
extremal cases lie the physical realistic cases of systems in
contact with \emph{finite heat baths}, \emph{where both energy and
temperature are allowed to fluctuate}. Therefore we find it
reasonable to expect such systems to obey Tsallis statistics of
some order $q$, where $q$ accounts for the specific heat of the
bath. The idea of finite heat bath is not new in the context of
non-extensive thermodynamics. It was first proposed in
\cite{Plast94}, and then further developed in \cite{Almeida01},
although the limiting case of null heat capacity was never
investigated before. In particular Almeida \cite{Almeida01}
considered an isolated system of total energy $a$ composed of two
non interacting subsystems: the system of interest (labelled by
the subscript $1$) and its complement, the bath (labelled by $2$).
The total Hamiltonian splits in the sum of the two sub-system
hamiltonians, $H = H_1+H_2$. Using the structure function of the
total system ($\Omega$), and that of the bath ($\Omega_2$), one
can express the distribution law for the component $1$, in its
phase space as \cite{Khinchin}:
\begin{equation}\label{} p_1(H_1) =
\frac{\Omega_2(a-H_1)}{\Omega(a)}.
\end{equation}
By defining the inverse temperature of the bath as $\beta_2 \doteq
\frac{\Omega'_2}{\Omega_2}$, Almeida proved that the bath specific
heat ($C_V^{-1} \doteq \frac{\partial}{\partial
E_2}\frac{1}{\beta_2}$) is given by the expression $C_V =
\frac{1}{1-q}$, if and only if:
\begin{equation}\label{eq:p1-ord}
    \frac{\Omega_2(a-H_1)}{\Omega_2(a)} = e_q(-\beta_2(a)H_1)
\end{equation}
namely, if and only if $p_1(H_1)\propto e_q(-\beta_2(a)H_1)$.
Although this theorem is in line with our interpretation based on
finite heat baths, it leads to a form of Tsallis distribution
expressed in terms of $e_q$ rather than the form investigated here
expressed in terms of $e'_q$. The latter would correspond to the
\emph{escort} version of the former. As the reader can easily
notice the microcanonical distribution is not a special case of
the $e_q$-type distribution. At this point we notice that the
definition $\beta_2 \doteq \frac{\Omega'_2}{\Omega_2}$ adopted by
Almeida is not consistent with the microcanonical equipartition
prescription (\ref{eq:equiTeo-microcanonical}). If one adopts the
correct definition
\begin{equation}\label{}
    \beta_2 \doteq \frac{\Omega_2}{\Phi_2},
\end{equation}
of bath's temperature, namely if one replaces $\Omega_2$ with
$\Phi_2$ then Almeida's theorem would read: $C_V =\frac{1}{1-q}
\Leftrightarrow \frac{\Phi_2(a-H_1)}{\Phi_2(a)} =
e_q(-\beta_2(a)H_1)$. By taking the derivative of the latter with
respect to $H_1$, we obtain the following:
\begin{theo}
\begin{equation}\label{eq:p1-nor}
   \frac{\Omega_2(a-H_1)}{\Omega_2(a)} =
    e'_q(-\beta_2(a)H_1)
\end{equation}
if and only if
\begin{equation}\label{eq:CV}
   C_V \doteq \left(\frac{\partial}{\partial
E_2}\frac{1}{\beta_2}\right)^{-1}=\frac{1}{1-q}.
\end{equation}
\end{theo}
Now we notice that $\beta_2(a)$ is the inverse physical
temperature that the bath would have if it were isolated and its
energy were $a$. Thanks to Eq. (\ref{eq:CV}) such temperature is
given by $\frac{1}{\beta_2(a)} = (1-q)a$. Instead, in the physical
situation under study, the bath is in contact with system $1$ and
its average energy $U_2$ is smaller than $a$. The actual
temperature in the composite system may be expressed by
$\frac{1}{\beta}=(1-q)U_2$. Therefore $\beta(a)=\beta U_2/a$,
hence Eq. (\ref{eq:p1-nor}) may be rewritten as $p_1(H_1) \propto
\left[ 1 + \left(\frac{a}{U_2}-1\right)- (1-q)\beta H_1
\right]^{\frac{q}{1-q}}$. Using the relation $a=U_1+U_2$ then
$p_1$ would read exactly as the Tsallis distribution of Eq.
(\ref{eq:Tsallis-ensemble}).

In sum, the theorem says that if the heat capacity of the bath is
$C_V = \frac{1}{1-q}$, then the component $1$ obeys Tsallis
distribution law of Eq. (\ref{eq:Tsallis-ensemble}) of index $q$.
This theorem is consistent with our physical interpretation
according to which $q$ should account for the finiteness of the
bath specific heat. In particular it reproduces well the two
limiting cases: if $q \rightarrow 1$ then $C_V$ goes to infinity,
namely we are in the case of infinite bath, and accordingly we get
the canonical ensemble. If $q \rightarrow -\infty$ then $C_V
\rightarrow 0$, namely we are in the isolated case, and
accordingly we get the microcanonical ensemble. Thanks to the
theorem above and the nonextensive equipartition theorem
($T=\frac{1}{\beta}$), we are finally in the position to write the
distribution law for system 1 in terms of physical quantities as:
\begin{equation}\label{}
    \rho_{C_V}(\textbf{z};U,V) = \frac{\left[ 1 -
\frac{(H(\textbf{z};V)-U)}{C_V T}\right]^{C_V-1}}{N_{C_V}(U,V)},
\end{equation}
where for simplicity we dropped the subscript 1. Accordingly its
entropy would read:
\begin{equation}\label{}
    S_{C_V}(U,V) = \log
     \int d\textbf{z} \left[ 1 -
\frac{(H(\textbf{z};V)-U)}{C_V T}\right]^{C_V}
\end{equation}

In conclusion the main novelty of this work is that of pointing
out that the microcanonical ensemble is a limiting case of Tsallis
ensemble. This can be seen if the escort version of Tsallis
distribution (namely the $e'_q$-type), which has been proved to be
an exact orthode \cite{Campisi06}, is adopted. By slightly
modifying Almeida's theorem \cite{Almeida01}, such statistics can
be proved to arise when the heat bath is finite. This, in turn,
allows to interpret the limit $q \rightarrow -\infty$ as the
physical situation in which the bath heat capacity tends to zero.

Within the proposed approach Tsallis statistics arise naturally as
a \emph{hybrid} statistics, where the qualifier \emph{hybrid} may
be understood at least in three different senses: (a) a
qualitative one, via the $\beta \leftrightarrow U$ \emph{duality};
(b) a mathematical one, via the fact that one can go from the
microcanonical ensemble to the canonical one and vice-versa
through a continuous family of Tsallis ensembles ranging from
$q=-\infty$ to $q=1$; and (c) a physical one, through the fact
that such statistics apply in the case of finite baths where
neither the energy nor the temperature are fixed but both are
allowed to fluctuate. Microcanonical and canonical cases are the
two ideal limiting cases of infinite and absent bath, where,
accordingly, the \emph{duality} of possible parameterizations is
lost. It is worth noticing that the idea of fluctuating
temperature is in agreement with Beck and Cohen's account of
nonextensive thermodynamics based on superstatistics
\cite{Beck03}. In this sense the Tsallis ensemble seems to have
the very nice feature of being able to describe a typical
out-of-equilibrium situation (fluctuating temperature) while
retaining the formal structure of equilibrium thermodynamics based
on average quantities (orthodicity).

\acknowledgments{Fruitful discussions with D. H. Kobe and G. B.
Bagci are gratefully acknowledged.}

\bibliography{thebibliography}

\newpage


\end{document}